\documentclass[aps,showpacs,preprintnumbers,amsmath, amssymb]{revtex4}

\usepackage{float}
\usepackage{graphics,epsfig}
\usepackage{graphicx}
\usepackage{dcolumn}
\usepackage{bm}

\begin{document}
\baselineskip=0.8 cm

\title{{\bf Scalar field condensation behaviors around reflecting shells in Anti-de Sitter spacetimes}}
\author{Yan Peng$^{1}$\footnote{yanpengphy@163.com},Bin Wang$^{2,3}$\footnote{wang\underline{~}b@sjtu.edu.cn}, Yunqi Liu$^{2}$\footnote{liuyunqi@hust.edu.cn}}
\affiliation{\\$^{1}$ School of Mathematical Sciences, Qufu Normal University, Qufu, Shandong 273165, China}
\affiliation{\\$^{2}$ Center for Gravitation and Cosmology, College of Physical Science
and Technology, Yangzhou University, Yangzhou 225009, China}
\affiliation{\\$^{3}$ Center of Astronomy and Astrophysics, Department of Physics and Astronomy,
Shanghai Jiao Tong University, Shanghai 200240, China}

\vspace*{0.2cm}
\begin{abstract}
\baselineskip=0.6 cm
\begin{center}
{\bf Abstract}
\end{center}

We study scalar condensations around asymptotically Anti-de Sitter(AdS) regular reflecting shells.
We show that the charged scalar field can condense around charged reflecting shells with
both Dirichlet and Neumann boundary conditions. In particular, the radii of the
asymptotically AdS hairy shells are discrete, which is similar to cases in asymptotically flat spacetimes.
We also provide upper bounds for the radii of the hairy Dirichlet reflecting shells and above the bound,
the scalar field cannot condense around the shell.

\end{abstract}

\maketitle
\newpage
\vspace*{0.2cm}

\section{Introduction}

Black holes in classical theories describe compact spacetime regions surrounded by event horizons,
which irreversibly absorb matter and radiation fields.
It was believed that the absorption at the horizon leads to the famous
no scalar hair theorem of black holes and no hair behaviors are restricted to spacetimes with horizons,
for reviews please refer to \cite{Bekenstein-1,CAR} and some other discussions can be found in \cite{Bekenstein}-\cite{Brihaye}.
The relation between the no hair theorem and the stability of perturbations around black hole spacetime was recently discussed in \cite{Hod-5,Hod-6}.

Whether no hair theorem can exist in regular gravity systems is a question to be answered.
Recently it was shown that no hair behavior can appear in the background of
asymptotically flat horizonless neutral compact reflecting star \cite{Hod-3}.
Moreover it was found that in asymptotically flat neutral compact reflecting stars,
massless scalar fields nonminimally coupled to gravity cannot exist \cite{Hod-4}.
The discussion was further generalized to spacetimes with a positive cosmological constant, where it was proved that  massive scalar,
vector and tensor hairs all die out outside the surface of compact reflecting stars, in consistent with the spirit of the no hair theorem \cite{Bhattacharjee}.

However, there are also some counter-examples  to challenge the no hair theorem obtained in regular gravity systems. It was found that charged scalar fields can condense around a charged reflecting shell, where the shell charge and mass can be neglected compared to the shell radius \cite{Hod-7,Hod-8}.
In addition, it was observed that the reflecting star can also support massive scalar fields \cite{Hod-9,YP2018}.
Moreover, around rapid spin compact reflecting stars  scalar condensations were observed \cite{Hod-10}.  Thus comparing to black hole spacetimes, it seems that no hair theorem does not generally hold in regular gravity systems.

Most available discussions on no hair theorem in horizonless spacetimes
were carried out in asymptotically flat and asymptotically dS backgrounds.
It is of great interest to extend the discussion to
asymptotically AdS regular backgrounds. The first attempt can be found in \cite{Bhattacharjee}.
With the interest of the AdS/CFT correspondence \cite{HS1,HS2,HS3,HS4},
the AdS hairy configurations have attracted a lot of attentions
and there are many discussions on the holographic condensation of scalar hair
in AdS spacetimes \cite{HS5}-\cite{HS20}.
Besides the holographic condensation, there is another way to construct hairy configurations, which can be prepared by enclosing the system in a scalar reflecting box \cite{box1}-\cite{box9}.
Similar to the box condition, it was found that there is an infinite potential
wall at the AdS boundary to confine the scalar field \cite{YDW}.
Considering the AdS boundary can confine the scalar field and make it easy to condense,
it is interesting to examine whether
no scalar hair theorem can exist in the asymptotically AdS horizonless spacetime.
It is intriguing to construct a regular hairy configuration in
the asymptotically AdS gravity.

The paper is organized as follows:  In section II,
we will introduce the system composed of a charged reflecting shell and a scalar field in the AdS gravity.
We emphasize here that the solution constructed here is a probe solution and that the shell is a scalar
shell. The background geometry is the AdS Schwarschild black hole. In the following section, we will discuss upper bounds on the radius of the scalar hairy Dirichlet reflecting shell. We will further obtain solutions of hairy Dirichlet reflecting shells and explore influences of
parameters on the largest hairy shell radii. We will generalize the discussion to the hairy Neumann reflecting shell.
Our summary and discussion can be found in the last section.

\section{Equations of motion and boundary conditions}

We consider the system of a scalar field coupled to the Maxwell field around a reflecting shell
in the four dimensional asymptotically AdS gravity.
We define the radial coordinate $r=r_{s}$ as the radius of the reflecting shell.
And the corresponding Lagrange density is given by
\begin{eqnarray}\label{lagrange-1}
\mathcal{L}=-\frac{1}{4}F^{MN}F_{MN}-|\nabla_{\mu} \psi-q A_{\mu}\psi|^{2}-m^{2}\psi^{2},
\end{eqnarray}
where q and $m$ are the charge and mass of the scalar field $\psi(r)$ respectively.
And $A_{\mu}$ stands for the ordinary Maxwell field.

Using the Schwarzschild coordinates, the line element of the
planar symmetric shell can be expressed in the form \cite{HS5,Chandrasekhar}
\begin{eqnarray}\label{AdSBH}
ds^{2}&=&-f(r)dt^{2}+\frac{dr^{2}}{f(r)}+r^{2}(dx^{2}+dy^{2}).
\end{eqnarray}
where $f(r)=\frac{r^2}{L^2}-\frac{2M}{r}+\frac{Q^2}{r^2}$ with $M$ and $Q$ as the mass and charge of the shell respectively.
And L is defined as the AdS radius.

For simplicity, we study the scalar field with only radial dependence in the form $\psi=\psi(r)$
and the Maxwell field is $A_{\nu}=-\delta_{\nu}^{0}Q/r$.
In this paper, we only concentrate on the case of $r_{s}\gg M,Q$ as assumed in \cite{Hod-7,Hod-8}.
It was found that properties in the $r_{s}\gg M,Q$ limit are qualitatively
the same as cases with nonzero mass and charge \cite{Hod-9,YP2018}.
From above assumptions, we obtain equations of motion as
\begin{eqnarray}\label{BHg}
\psi''+\frac{4}{r}\psi'+(\frac{q^2Q^2L^4}{r^6}-\frac{m^2L^2}{r^2})\psi=0.
\end{eqnarray}

We can simply set $L=1$ in equation (3) with the symmetry \cite{Steven}
\begin{equation}
L\rightarrow \alpha L,~~r\rightarrow \alpha r,~~q\rightarrow q/\alpha,~~t\rightarrow \alpha t,~~Q\rightarrow \alpha Q,~~m\rightarrow m/\alpha.
\end{equation}

Near the AdS boundary $r\rightarrow\infty$, the asymptotic behaviors of the scalar fields are
\begin{eqnarray}\label{InfBH}
&&\psi=\frac{\psi_{-}}{r^{\lambda_{-}}}+\frac{\psi_{+}}{r^{\lambda_{+}}}+\cdots,
\end{eqnarray}
with $\lambda_{\pm}=(3\pm \sqrt{9+4m^2})/2$.
For $m^2>0$, only the second scalar operator $\psi_{+}$ is normalizable
and we fix $\psi_{-}=0$. Then the boundary condition at the infinity is
\begin{equation}
\psi(\infty)=0.
\end{equation}

In addition, we impose Dirichlet or Neumann reflecting boundary conditions for the scalar field
at the surface of the shell. And the two types of surface boundary conditions are

 \begin{equation}
\left\{
\begin{array}{l}
\psi(r_{s})=0~~~~~~~~~~~Dirichlet~~~B.C;\\
\\
d\psi(r_{s})/dr=0~~~~~Neumann~~~B.C.\\
\end{array}
\right.
\end{equation}

\section{Scalar field condensations around charged reflecting shells}

\subsection{Upper bounds for radii of scalar hairy Dirichlet reflecting shells}

Defining the new radial function $\tilde{\psi}=\sqrt{r}\psi$,
one obtains the differential equation
\begin{eqnarray}\label{BHg}
r^2\tilde{\psi}''+3r\tilde{\psi}'+(-\frac{5}{4}+\frac{q^2Q^2L^4}{r^4}-m^2L^2)\tilde{\psi}=0.
\end{eqnarray}

According to the boundary conditions (6) and (7), one deduces that
\begin{eqnarray}\label{InfBH}
&&\tilde{\psi}(r_{s})=0,~~~~~~~~~\tilde{\psi}(\infty)=0.
\end{eqnarray}

The function $\tilde{\psi}$ must have (at least) one extremum point $r=r_{peak}$
between the surface $r_{s}$ of the reflecting shell and the AdS boundary $r_{b}=\infty$.
At this extremum point, the scalar field is characterized by
\begin{eqnarray}\label{InfBH}
\{ \tilde{\psi}'=0~~~~and~~~~\tilde{\psi} \tilde{\psi}''\leqslant0\}~~~~for~~~~r=r_{peak}.
\end{eqnarray}

With the relations (8) and (10), we obtain the inequality
\begin{eqnarray}\label{BHg}
-\frac{5}{4}+\frac{q^2Q^2}{(\frac{r}{L})^4}-m^2L^2\geqslant0~~~for~~~r=r_{peak}.
\end{eqnarray}

Then we arrive at an upper bound for the radius of the scalar hairy shell
\begin{eqnarray}\label{BHg}
\frac{r_{s}}{L}\leqslant \frac{r_{peak}}{L} \leqslant\sqrt[4]{\frac{q^2Q^2}{m^2L^2+\frac{5}{4}}}
\end{eqnarray}

According to the symmetry (4), this upper bound can be expressed with dimensionless quantities as
\begin{eqnarray}\label{BHg}
m r_{s}\leqslant m r_{peak} \leqslant mL\sqrt[4]{\frac{q^2Q^2}{m^2L^2+\frac{5}{4}}}
\end{eqnarray}

From the upper bound, there is $m r_{s}=0$ for $Q=0$. So there is no hairy neutral shell or neutral shells
cannot support scalar fields. In another case of $q=0$,
again we have $m r_{s}=0$ or there is no scalar hair behaviors for neutral scalar fields.

\subsection{Construction of charged scalar hairy Dirichlet reflecting shells}

In this part, we give an analytical treatment of the AdS hairy reflecting shell
in the limit of $M,Q\ll r_{s}$.
With the symmetry (4), we also fix $L=1$ in the calculation.
The general solutions of the radial differential equation (8) can be
expressed in terms of the Bessel functions \cite{Abramowitz}.
In the case of $1-\nu \neq 0,-1,-2,-3,\cdots$, the solutions are
\begin{eqnarray}\label{BHg}
\tilde{\psi}(r)=A_{0}\cdot \frac{1}{r^{3/2}}
  J_{\nu}(\frac{qQ}{2r^2})+B_{0}\cdot \frac{1}{r^{3/2}}
  J_{-\nu}(\frac{qQ}{2r^2})
\end{eqnarray}
and for $1-\nu=n= 0,-1,-2,-3,\cdots$, the solutions are
\begin{eqnarray}\label{BHg}
\tilde{\psi}(r)=A_{1}\cdot \frac{1}{r^{3/2}}
  J_{\nu}(\frac{qQ}{2r^2})+B_{1}\cdot \frac{1}{r^{3/2}}
  Y_{\nu}(\frac{qQ}{2r^2}),
\end{eqnarray}
where $\nu=\frac{\sqrt{9+4 m^2}}{4}$ and $A_{i}$,$B_{i}$ are constants.

The large r behavior of the radial solution (14) and (15) is given by
\begin{eqnarray}\label{BHg}
\tilde{\psi}(r)\propto A_{i}\cdot \frac{1}{r^{(3+ \sqrt{9+4 m^2})/2}}+B_{i}\cdot \frac{1}{r^{(3- \sqrt{9+4 m^2})/2}}.
\end{eqnarray}

Considering the boundary condition (9), one deduces that the coefficient $B_{i}$
must vanish: $B_{i}=0$.
We therefore conclude that the bound-state configurations of the charged massive scalar fields in the background of
the charged shell are characterized by the radial eigenfunction
\begin{eqnarray}\label{BHg}
\tilde{\psi}(r)\propto \frac{1}{r^{3/2}}
  J_{\nu}(\frac{qQ}{2r^2}).
\end{eqnarray}

With the boundary condition (9) and equation (17), one finds the characteristic resonance equation
\begin{eqnarray}\label{BHg}
J_{\nu}(\frac{qQ}{2r^2})=0
\end{eqnarray}
for the composed charged shell and charged massive scalar field configurations.
Interestingly, as we shall show below, the resonance condition (18) determines the discrete shell radius, which can
support the charged massive scalar field.
The analytically derived resonance condition (18) can be solved numerically.
For given set of parameters, there exists a discrete set of shell radius
\begin{eqnarray}\label{BHg}
\cdots<mR_{2}<mR_{1}<mR_{0}=mR_{max}\leqslant mL\sqrt[4]{\frac{q^2Q^2}{m^2L^2+\frac{5}{4}}}
\end{eqnarray}
which can support the static charged massive scalar field.
For given set of parameters, we can search for the largest radius $mR_{max}$
below the upper bound.
From the equation (18), it is clearly that larger $qQ$ corresponds to a larger hairy shell radius
and we can define a new coordinate $\frac{r_{s}}{\sqrt{qQ}}$.
In Table I, we display the largest
radius of the hairy charged shell $mR_{max}$ for various parameters $m^{2}L^2$ and $qQ$
with dimensionless quantities according to the symmetry (4).
In Fig. 1, we further show behaviors of $\frac{mR_{max}}{\sqrt{qQ}}$ with various $m^2L^2$ in blue line.
It shows that larger scalar field mass leads to a larger hairy shell radius.
We fit the data and arrive at an approximate relation $mR_{max}\approx mL(0.3784-0.0097m^2L^2)\sqrt{qQ}$
as shown in red line of Fig. 1.
We can see from the picture that the red line almost coincides with the blue line.

\renewcommand\arraystretch{1.7}
\begin{table} [h]
\centering
\caption{The largest radius $mR_{max}$ together with various $m^2L^2$ and qQ}
\label{address}
\begin{tabular}{|>{}c|>{}c|>{}c|>{}c|>{}c|>{}c|}
\hline
$~~~~~~~m^{2}L^2~~~~~~~$ & ~~~~~~~0~~~~~~~ & ~~~~~~~1/2~~~~~~~& ~~~~~~~1~~~~~~~& ~~~~~~~3/2~~~~~~~& ~~~~~~~2~~~~~~~\\
\hline
$~~~~~~~\frac{mR_{max}}{\sqrt{qQ}}~~~~~~~$ & ~~~~~~~0~~~~~~~ & ~~~~~~~0.2638~~~~~~~& ~~~~~~~0.3682~~~~~~~& ~~~~~~~0.4450~~~~~~~& ~~~~~~~0.5088~~~~~~~\\
\hline
\end{tabular}
\end{table}

\begin{figure}[h]
\includegraphics[width=180pt]{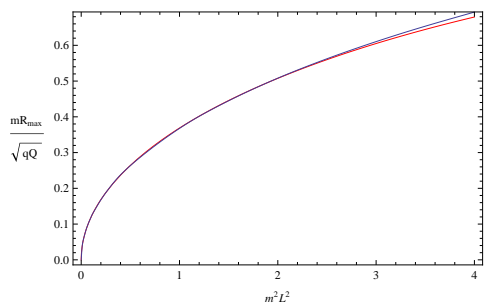}\
\caption{\label{EEntropySoliton} (Color online) We show the largest hairy shell radius $\frac{mR_{max}}{\sqrt{qQ}}$ as a function of
the scalar field mass $m^2L^2$ in blue line. And the red line is with the fitting formula: $\frac{mR_{max}}{\sqrt{qQ}}\approx mL(0.3784-0.0097m^2L^2)$.}
\end{figure}

\subsection{Scalar condensations around a charged Neumann reflecting shell}

In this part, we study scalar field condensations supported by
the Neumann reflecting shell. From the relation (17), we have $\psi=0$ for $qQ=0$.
It means the no scalar hair behaviors also exist for the
case of $qQ=0$ with Neumann reflecting shells.
In contrast, we will show in the following that there is scalar field configurations
supported by Neumann reflecting shells in the case of $qQ\neq 0$.

According to Neumann boundary conditions (7),
the hairy shell radius is characterized by the resonance equation
\begin{eqnarray}\label{BHg}
\frac{d}{dr}[\frac{1}{r^{3/2}}J_{\nu}(\frac{qQ}{2r^2})]\mid_{r=r_{s}}=0.
\end{eqnarray}

We solve the resonance condition (20) with numerical methods.
For given set of parameters, there exists a discrete set of shell radius
\begin{eqnarray}\label{BHg}
\cdots~<~m\tilde{R}_{2}~<~m\tilde{R}_{1}~<~m\tilde{R}_{0}~=~m\tilde{R}_{max}
\end{eqnarray}
which can support the static charged massive scalar field. In Table II, we display the largest
radius of the hairy charged shell $m\tilde{R}_{max}$ for different values of the scalar mass $m^{2}L^2$
and in Table III, we show the largest
radius of the hairy charged shell with various parameter $qQ$.
It can be easily seen that a larger scalar mass corresponds to a larger $m\tilde{R}_{max}$
and a larger $qQ$ also leads to a larger $m\tilde{R}_{max}$.

\renewcommand\arraystretch{1.5}
\begin{table} [h]
\centering
\caption{The largest radius $m\tilde{R}_{max}$ together with various $m^2L^2$ and qQ=1}
\label{address}
\begin{tabular}{|>{}c|>{}c|>{}c|>{}c|>{}c|>{}c|}
\hline
$~~~~~~~m^{2}L^2~~~~~~~$ & ~~~~~~~0~~~~~~~ & ~~~~~~~1/2~~~~~~~& ~~~~~~~1~~~~~~~& ~~~~~~~3/2~~~~~~~& ~~~~~~~2~~~~~~~\\
\hline
$~~~~~~~m\tilde{R}_{max}~~~~~~~$ & ~~~~~~~0~~~~~~~ & ~~~~~~~0.3452~~~~~~~& ~~~~~~~0.4788~~~~~~~& ~~~~~~~0.5764~~~~~~~& ~~~~~~~0.6553~~~~~~~\\
\hline
\end{tabular}
\end{table}

\renewcommand\arraystretch{1.7}
\begin{table} [h]
\centering
\caption{The largest radius $m\tilde{R}_{max}$ together with various $\sqrt{qQ}$ and $m^2L^2=1$}
\label{address}
\begin{tabular}{|>{}c|>{}c|>{}c|>{}c|>{}c|>{}c|}
\hline
$~~~~~~~\sqrt{qQ}~~~~~~~$ & ~~~~~~~0~~~~~~~ & ~~~~~~~1/2~~~~~~~& ~~~~~~~1~~~~~~~& ~~~~~~~3/2~~~~~~~& ~~~~~~~2~~~~~~~\\
\hline
$~~~~~~~m\tilde{R}_{max}~~~~~~~$ & ~~~~~~~0~~~~~~~ & ~~~~~~~0.2394~~~~~~~& ~~~~~~~0.4788~~~~~~~& ~~~~~~~0.7182~~~~~~~& ~~~~~~~0.9577~~~~~~~\\
\hline
\end{tabular}
\end{table}

We also show behaviors of $m\tilde{R}_{max}$ in Fig. 2.
We plot $m\tilde{R}_{max}$ as a function of $m^2L^2$ with $qQ=1$ in the left panel
and we show $m\tilde{R}_{max}$ as a function of $\sqrt{qQ}$ with $m^2L^2=1$ in the right panel.
We find that larger scalar field mass leads to a larger hairy shell radius
and larger $qQ$ also corresponds to a larger hairy shell radius.
And $m\tilde{R}_{max}$ is almost in linear with respect to $\sqrt{qQ}$ in the right panel of Fig. 2.
We fit the data and arrive at an approximate relation $m\tilde{R}_{max}\approx mL(0.4979-0.0178m^2L^2)\sqrt{qQ}$.
It can be seen that the red lines almost coincide with the blue lines in both panels.

\begin{figure}[h]
\includegraphics[width=165pt]{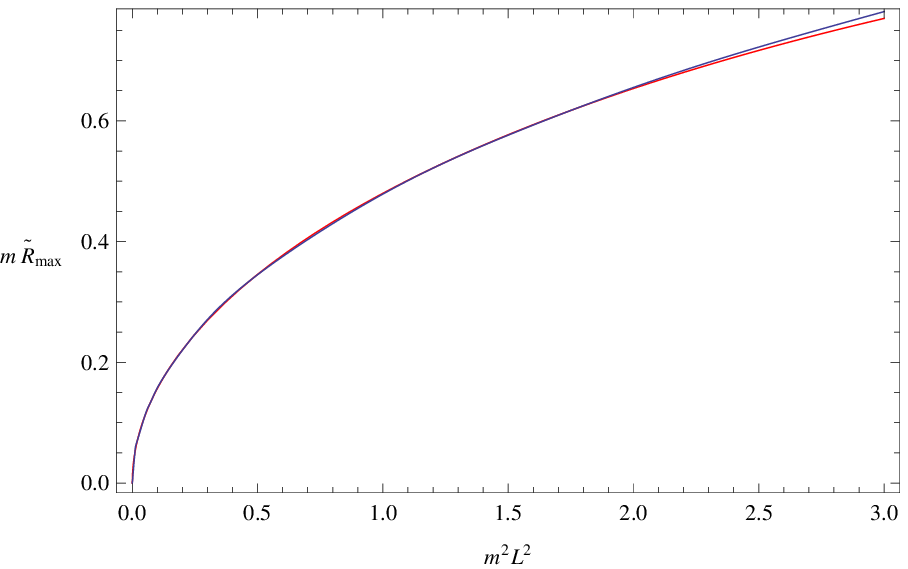}\
\includegraphics[width=165pt]{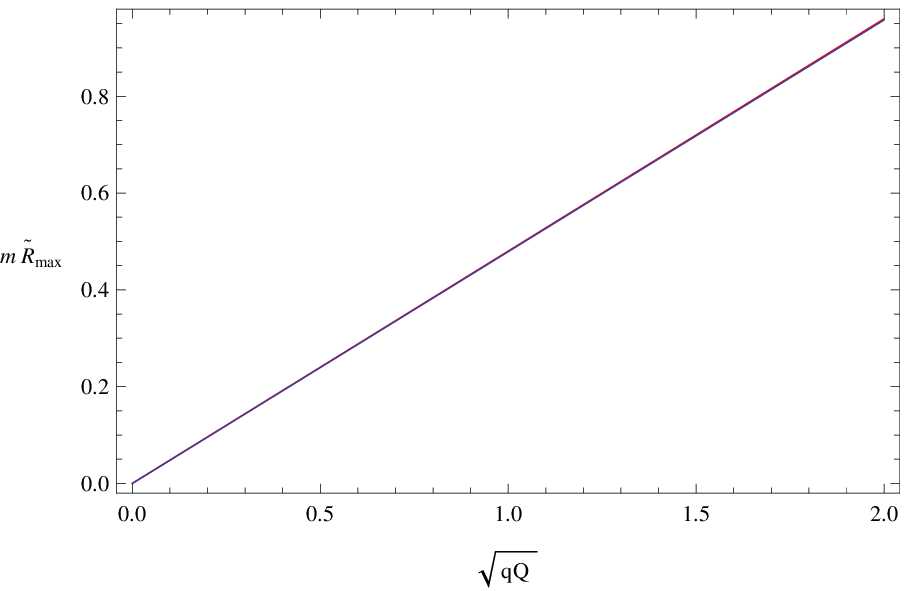}\
\caption{\label{EEntropySoliton} (Color online) In the left panel, we show the largest hairy shell radius $m\tilde{R}_{max}$ as a function of
the scalar field mass $m^2L^2$ with $qQ=1$ in blue lines and in the right panel, we represent $m\tilde{R}_{max}$ as a function of
$\sqrt{qQ}$ with $m^2L^2=1$ in blue curves. And the red curves in both panels correspond to the fitting formula $m\tilde{R}_{max}\approx mL(0.4979-0.0178m^2L^2)\sqrt{qQ}$.}
\end{figure}

From the approximate formulae and curves in the left panels of Fig. 1 and Fig. 2,
we can easily see that behaviors of largest radius are similar for different boundary conditions.
We further compare the largest radius with different boundary conditions in Fig. 3.
It can be seen that $m\tilde{R}_{max}$ is larger than $mR_{max}$ for fixed $m^2L^2$ and $qQ$.
This property is qualitatively the same as results in asymptotically flat spacetime \cite{Hod-9}.

\begin{figure}[h]
\includegraphics[width=180pt]{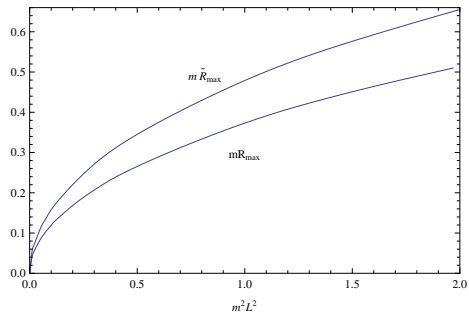}\
\caption{\label{EEntropySoliton} (Color online) We show the largest hairy shell radius as a function
of the scalar mass with $qQ=1$ and different boundary conditions. The top line represents
behaviors of $m\tilde{R}_{_{max}}$ with Neumann reflecting conditions
and the bottom line shows behaviors of $mR_{max}$ with Dirichlet reflecting conditions.}
\end{figure}

\section{Conclusions}

We studied condensations of static scalar fields around a regular reflecting shell in the AdS spacetime.
We considered both Dirichlet and Neumann reflecting boundary conditions for
the scalar field at the surface of the shell.
In the case of Dirichlet boundary conditions,
we provided upper bounds for the radius of the scalar hairy shell
as $m r_{s}\leqslant mL\sqrt[4]{\frac{q^2Q^2}{m^2L^2+\frac{5}{4}}}$.
This upper bound showed that the no scalar hair theorem exists in the case of $qQ=0$
corresponding to neutral scalar fields or neutral shells.
In other cases of $qQ\neq0$, we found that charged scalar fields can condense around the charged Dirichlet reflecting shell.
And the radius of the hairy shell is discrete similar to cases in the asymptotically flat space.
For each set of fixed parameters, there is a largest hairy shell radius $mR_{max}$.
We found that larger scalar field mass correspond to a larger $m R_{max}$
and larger qQ is also with a larger $m R_{max}$.
We further obtained an approximate relation $mR_{max}\approx mL(0.3784-0.0097m^2L^2)\sqrt{qQ}$.
Moreover, we constructed charged scalar field configurations supported
by charged Neumann reflecting shells and the hairy shell radii are also discrete.
We fitted the data and arrived at an approximate relation $m\tilde{R}_{max}\approx mL(0.4979-0.0178m^2L^2)\sqrt{qQ}$
with $m\tilde{R}_{max}$ as largest radii of hairy Neumann reflecting shells.

\begin{acknowledgments}

We would like to thank the anonymous referee for the constructive suggestions to improve the manuscript.
Yan Peng was supported by the Shandong Provincial Natural Science
Foundation of China under Grant No. ZR2018QA008.
Bin Wang would like to acknowledge the support by National Basic Research Program of China (973 Program 2013CB834900)
and National Natural Science Foundation of China.
And Yunqi Liu acknowledge the support by the National Natural Science Foundation of China under Grant No. 11505066.

\end{acknowledgments}

\end{document}